\documentclass[fp,twocolumn]{jpsj3}
\usepackage{txfonts}
\usepackage{graphicx}
\usepackage{bm} % bold math

\title{Energy Scale Deformation on Regular Polyhedra}

\author{Takuya Eguchi$^{1}$, Satoshi Oga$^{2}$, Hosho Katsura$^{3,4,5}$, Andrej Gendiar$^{6}$, 
and Tomotoshi Nishino$^{1}$\thanks{nishino@kobe-u.ac.jp}}
\inst{$^1$Department of Physics, Graduate School of Science, Kobe University, Kobe 657-8501, Japan\\
$^2$The Institute for Solid State Physics, The University of Tokyo, Kashiwa, Chiba 277-8581, Japan\\
$^3$Department of Physics, The University of Tokyo, 
7-3-1 Hongo, Bunkyo-ku, Tokyo 113-0033, Japan\\
$^4$Institute for Physics of Intelligence, The University of Tokyo, 
7-3-1 Hongo, Bunkyo-ku, Tokyo 113-0033, Japan\\
$^5$Trans-scale Quantum Science Institute, University of Tokyo, 
Bunkyo-ku, Tokyo 113-0033, Japan\\
$^6$Institute of Physics, Slovak Academy of Sciences, Bratislava 845 11, Slovakia}

\abst{
A variant of energy scale deformation is considered for the $S = 1/2$ antiferromagnetic 
Heisenberg model on polyhedra. The deformation is induced by the perturbations to the 
uniform Hamiltonian, whose coefficients are determined by the bond coordinates. 
On the tetrahedral, octahedral, and cubic clusters, the perturbative terms do not affect 
the ground state of the uniform Hamiltonian when they are sufficiently small.  
On the other hand, for the icosahedral and dodecahedral clusters, it is numerically confirmed 
that the ground state of the uniform Hamiltonian is almost insensitive to the perturbations 
unless they lead to a discontinuous change in the ground state. The obtained results suggest 
the existence of a generalization of sine-square deformation in higher dimensions.
}

\begin{document}
\maketitle

\section{Introduction}

Uniformity in quantum states is one of the fundamental properties in condensed matter physics. 
On a regular lattice, the ground state of a quantum model is expected to be uniform when the 
Hamiltonian is translationally invariant unless spatial modulations are spontaneously stabilized. 
Occasionally, the ground state is uniform even when the Hamiltonian is not translationally 
invariant. For example, when the excitation gap is sufficiently large, the effect of slowly varying  
%-ed-% perturbations is negligible to the uniform ground state. 
perturbations on the uniform ground state is negligible. 

A class of non-uniform Hamiltonians, whose ground states are nearly uniform, is known in one 
dimension. Suppose that we have a one-dimensional lattice Hamiltonian
\begin{equation}
{\hat H} = \sum_{\ell}^{~} \, {\hat h}_{\ell, \ell+1}^{~} \, ,
\end{equation}
where ${\hat h}_{\ell, \ell+1}^{~}$ represents the nearest-neighbor interaction, whose 
magnitude does not depend on the site index $\ell$. In the case of the translationally 
invariant quantum Heisenberg spin chain, ${\hat h}_{\ell, \ell+1}^{~}$ is written as the 
exchange interaction $J \, {\hat {\bm S}}_{\ell}^{~}  \cdot {\hat {\bm S}}_{\ell + 1}^{~}$, 
where ${\hat {\bm S}}_{\ell}^{~}$ denotes the spin operator at site $\ell$, and $J$ is the 
interaction parameter. In what follows, we assume that the ground state is uniform, and the 
excitation gap is zero in the thermodynamic limit. Introducing a deformation function 
$f_{\ell}^{~}$, which varies slowly with respect to $\ell$, we can modify {\it the energy scale} 
of each bond and define the non-uniform Hamiltonian
\begin{equation}
{\hat H}_f^{~} = \sum_{\ell}^{~} \, f_{\ell}^{~} \, {\hat h}_{\ell, \ell+1}^{~} \, .
\end{equation}
When the function is exponential, i.e., $f_{\ell}^{~} = e^{\ell / \lambda}_{~}$,~\cite{Okunishi1} the 
ground-state of ${\hat H}_f^{~}$ is uniform in the bulk part of the system.~\cite{Okunishi2} Under 
this {\it exponential deformation}, the correlation length becomes finite, and increases with the 
deformation parameter $\lambda > 0$. A similar uniformity of the ground state has been 
observed for the hyperbolic deformation function 
$f_{\ell}^{~} = \cosh( \ell / \lambda )$.~\cite{Ueda0,Ueda1,Ueda2}

The specific form of deformation that we focus on in this article is the {\it sine-square deformation} 
(SSD).~\cite{Gendiar1,Gendiar2,Hikihara} Consider the $N$-site system whose Hamiltonian is 
written as
\begin{equation}
{\hat H}_{\rm SSD}^{~} = \sum_{\ell = 1}^{N} \, 2 \left[ \sin\frac{\ell \pi}{N}  \right]^2_{~} \, 
{\hat h}_{\ell, \ell+1}^{~} \, ,
\end{equation}
where we have used the labeling rule that identifies $\ell = N +1$ with $\ell = 1$. 
The prefactor of ${\hat h}_{N,1}^{~}$ is zero, and therefore there is no coupling 
between the ends $\ell = 1$ and $\ell = N$. 
%-ed-% Thus the system 
Thus, the system 
can be considered as the finite-size system of length $N$ with open and smooth 
%-ed-% boundary condition,~\cite{smooth1,smooth2} 
boundary conditions,~\cite{smooth1,smooth2} 
where the interaction strength decreases 
toward the both ends of the system. It was accidentally found that the ground state 
is uniform under the SSD when the free fermionic lattice model is 
considered.~\cite{Gendiar1,Gendiar2}
In the correlated systems, the uniformity under the SSD was numerically confirmed 
for the Kondo lattice model,~\cite{Shibata} the $S = 1/2$ Heisenberg spin chain,~\cite{Hikihara}
and the Hubbard model.~\cite{Daniska} Theoretical proof of complete uniformity is given
for the free fermionic lattice model.~\cite{Katsura,Okunishi3} It has been known that the
continuum limit of the SSD has a natural interpretation in terms of 
%-ed-% conformal field theory 
%-ed-%%-ed-% the conformal field theory 
conformal field theory 
(CFT).~\cite{Katsura2,Tada1,Ishibashi1,Ishibashi2,Okunishi4,Tamura,Wen1,Tada2,
Wen2,Wen3,Kishimoto,Zeze,MacC,Tada3,Santos,Fan1,Chen,Lapierre1,Liu,Caputa,Lapierre2,
Wen4,Ageev,Han,Andersen,Fan,Das} 
Generalizations of the SSD to two dimensions were considered on finite lattices 
with torus,~\cite{Maruyama} disk,~\cite{Hotta,Nishimoto,Asano} and 
tube geometries.~\cite{Nishimoto,Yonaga}

\begin{figure}
\begin{center}
\centering\includegraphics[width = 70 mm]{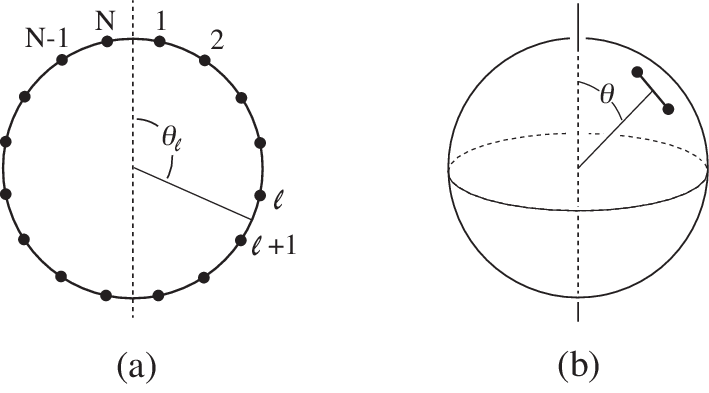}
\end{center}
\caption{(a) Angle $\theta_{\ell}^{~}$ contained in ${\hat H}_{\rm M}^{~}$ in Eq.~(6). 
(b) Angle $\theta$ for a bond 
%-ed-% which 
that 
connects lattice points on the sphere.}
\end{figure}

The deformation function in Eq.~(3) can be written as
\begin{equation}
f_{\ell}^{~} = 
2 \left[ \sin\frac{\ell \pi}{N}  \right]^2_{~} = 
1 - \cos\frac{2\ell \pi}{N} \, .
\end{equation}
Accordingly, we can decompose ${\hat H}_{\rm SSD}^{~}$ into a sum of the uniform part
\begin{equation}
{\hat H}_0^{~} = \sum_{\ell = 1}^{N} \, {\hat h}_{\ell, \ell+1}^{~}
\end{equation}
and the modulated part
\begin{equation}
{\hat H}_{\rm M}^{~} = 
- \sum_{\ell = 1}^{N} \, \cos\frac{2 \ell \pi}{N} \,\, {\hat h}_{\ell, \ell+1}^{~} \, .
\end{equation}
A simple geometrical interpretation is possible for the angle $\theta_{\ell}^{~} = 2 \ell \pi / N$
in Eq.~(6), as shown in 
%-ed-% Fig.~1 (a). 
Fig.~1(a). 
When the lattice sites are 
%-ed-% equidistantly put
located equidistantly 
on the circle, 
$\theta_{\ell}^{~}$ corresponds to the angle between the midpoint of the bond $( N,1 )$ 
and that of the bond $( \ell, \ell+1 )$ measured from the center of the circle. In the case of 
the free-fermion hopping model on the lattice, it is straightforward to show that the ground 
state $| \, \psi_0^{~} \rangle$ of the uniform part ${\hat H}_0^{~}$ is an eigenstate of 
the modulated part ${\hat H}_{\rm M}^{~}$ with eigenvalue zero. It has been analytically 
shown that the generalized Hamiltonian 
\begin{equation}
{\hat H}( \gamma ) = {\hat H}_0^{~} + \gamma \, {\hat H}_{\rm M}^{~}
\end{equation}
shares the same ground state $| \, \psi_0^{~} \rangle$ within the range 
$| \gamma | \le 1$.~\cite{Katsura, Maruyama} 

%%%%%-begin-%%%%%
From the construction of ${\hat H}_{\rm M}^{~}$ in Eq.~(6), which is related to the $N$-sided 
regular polygon in 
%-ed-% Fig.~1 (a),
Fig.~1(a),
 it is possible to state that ${\hat H}_{\rm M}^{~}$ corresponds to 
the most slowly varying sinusoidally modulated function on the finite lattice. This geometric 
observation suggests a new type of two-dimensional generalization of the SSD. Let us 
imagine discrete lattices drawn on a sphere. The possible candidates are finite lattices 
in the shape of regular polyhedra, which are tetrahedron ($N = 4$), octahedron ($N = 6$), 
cube ($N = 8$), icosahedron ($N = 12$), and dodecahedron ($N = 20$). Ground-state 
properties of the $S = 1/2$ antiferromagnetic Heisenberg model on these lattices have been 
known for the uniform case ${\hat H}_0^{(N)}$.~\cite{Modine,Konstantinidis0,Konstantinidis1,
Konstantinidis2,Konstantinidis3,Karlova,Konstantinidis4,Tabrizi} 
Recall that the most slowly varying function on the unit sphere is the spherical harmonic
function $Y_1^0 \propto \cos\theta$, where $\theta$ represents the angle from a fixed axis,  
and $\cos\theta$ represents the coordinate component along the axis. 
%-ed-% Figure 1 (b) shows 
Figure 1(b) shows 
the angle $\theta$ for a bond 
%-ed-% which
that connects lattice points on the sphere. 
We thus introduce the modulated part ${\hat H}_M^{(N)}$ that is the sum of the 
non-uniform nearest-neighbor interactions whose coefficients are specified by a linear 
function of coordinates of each bond. In this article, we 
%%%%%-end-%%%%%
examine the effect of ${\hat H}_M^{(N)}$ by means of obtaining the ground state of the 
combined Hamiltonian ${\hat H}^{(N)}_{~} = {\hat H}_0^{(N)} + c \, {\hat H}_M^{(N)}$. 
For $N = 4$, $6$, and $8$, it is confirmed that the ground state of ${\hat H}_0^{(N)}$ is
also the eigenstate of ${\hat H}_M^{(N)}$ with eigenvalue zero, and thus the ground state of 
${\hat H}^{(N)}_{~}$ is independent of $c$ when $|c|$ is relatively small. 
For $N = 12$ and $20$, the ground state of ${\hat H}_M^{(N)}$ depends on $c$, 
but the observed $c$-dependences are very weak. 

The structure of this article is as follows. 
In the next section, we consider the tetrahedral cluster, which can be treated analytically.
In 
%-ed-% \S 3, 
Sect. 3, 
the octahedral and 
%-ed-% the 
cubic clusters are examined. 
In 
%-ed-% \S 4, 
Sect. 4, 
the icosahedral and 
%-ed-% the 
dodecahedral clusters are examined. 
In these cases, the modulation changes the ground state, but the effect is very weak.
Conclusions are summarized in the last section. We discuss possible generalizations of 
the SSD in higher dimensions.

\section{Energy Scale Deformation on the Tetrahedral Cluster}

Consider the $S = 1/2$ antiferromagnetic Heisenberg model on finite lattices 
in the shape of regular polyhedra. Throughout this article, we assume 
only the nearest-neighbor interactions.
%-ed-% only. 
We set the interaction parameter $J$ to unity, and
thus the interaction between the neighboring sites $i$ and $j$ is simply expressed as 
${\hat h}_{i,j}^{~} = {\hat {\bm S}}_i^{~} \cdot {\hat {\bm S}}_j^{~}$. 

\begin{figure}
\begin{center}
\centering\includegraphics[width = 51 mm]{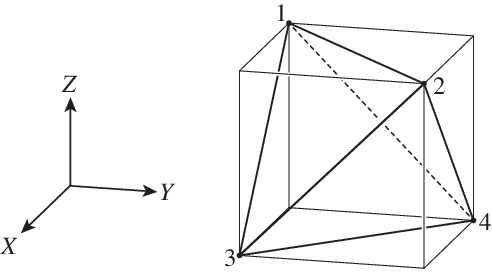}
\end{center}
\caption{Positions of sites on the tetrahedral cluster. }
\end{figure}

To 
%-ed-% get used to 
become familiar with 
polyhedral geometries, we start with the tetrahedral cluster 
shown in Fig.~2, which is drawn inside the cube. The coordinates of 
%-ed-% $\ell$-th 
the $\ell$-th 
site $\ell: ( x, y, z )$ are fixed to
\begin{align}
& 1:( - 1 , - 1 , ~~\,1 ) \, , ~~~~ 2:( ~~1, ~~1, ~~\,1 ) \, , \nonumber\\
& 3:( ~~1, - 1, - 1 ) \, , ~~~~ 4:( - 1, ~~1, - 1 ) \, ,
\end{align}
where we set the origin at the center of the cube. The uniform Hamiltonian on 
the cluster is represented as
\begin{eqnarray}
{\hat H}_0^{(4)} 
\!\!\!\! &=& \!\!\!\! 
{\hat {\bm S}}_2^{~} \cdot {\hat {\bm S}}_3^{~} + 
{\hat {\bm S}}_1^{~} \cdot {\hat {\bm S}}_4^{~} + 
{\hat {\bm S}}_2^{~} \cdot {\hat {\bm S}}_4^{~} + 
{\hat {\bm S}}_1^{~} \cdot {\hat {\bm S}}_3^{~} + 
{\hat {\bm S}}_1^{~} \cdot {\hat {\bm S}}_2^{~} + 
{\hat {\bm S}}_3^{~} \cdot {\hat {\bm S}}_4^{~} \nonumber\\
 \!\!\!\! &=& \!\!\!\! 
\frac{1}{2}\Bigl( 
{\hat {\bm S}}_1^{~} + {\hat {\bm S}}_2^{~} + 
{\hat {\bm S}}_3^{~} + {\hat {\bm S}}_4^{~} \Bigr)^2_{~} - \frac{3}{2}
 \, .
\label{eq:tetrahedron}
\end{eqnarray}
We have explicitly shown the number of sites $N = 4$ in the notation ${\hat H}_0^{(4)}$.
This is an example of the 
%-ed-% Majumdar-Ghosh 
Majumdar--Ghosh 
lattice.~\cite{M-G} 
Equation (\ref{eq:tetrahedron}) clearly shows that the model ${\hat H}_0^{(4)}$ has two 
degenerate ground states with total spin $0$, each of which can be written as a product of 
singlet pairs. The other eigenstates of ${\hat H}_0^{(4)}$ have a nonzero total spin. 

In order to introduce spatial modulations to the interactions, we focus on the coordinate of the 
midpoint of each bond. For example, they are $(0, 0, 1)$ and $(0, 0, -1)$, respectively, 
for the bonds $( 1, 2 )$ and $( 3, 4 )$. If we use the $Z$-component of the midpoint 
coordinate as the prefactor to the corresponding pairwise interaction, 
we obtain the following modulated 
%-ed-% part 
part:
\begin{equation}
{\hat H}_Z^{(4)} \, =  \,
{\hat {\bm S}}_1^{~} \cdot {\hat {\bm S}}_2^{~} - 
{\hat {\bm S}}_3^{~} \cdot {\hat {\bm S}}_4^{~} \, .
\end{equation}
It can be easily verified that ${\hat H}_Z^{(4)}$ commutes with ${\hat H}_0^{(4)}$. 
Therefore, the combined Hamiltonian
\begin{eqnarray}
{\hat H}_{~}^{(4)}( \gamma ) \!\!\!\! &=& \!\!\!\!  {\hat H}_0^{(4)} + \gamma \, {\hat H}_Z^{(4)} 
\nonumber\\
\!\!\!\! &=& \!\!\!\! 
{\hat {\bm S}}_2^{~} \cdot {\hat {\bm S}}_3^{~} + 
{\hat {\bm S}}_1^{~} \cdot {\hat {\bm S}}_4^{~} + 
{\hat {\bm S}}_2^{~} \cdot {\hat {\bm S}}_4^{~} + 
{\hat {\bm S}}_1^{~} \cdot {\hat {\bm S}}_3^{~} 
\nonumber\\
&&  \!\!\!\! + \, 
( 1 + \gamma ) \, {\hat {\bm S}}_1^{~} \cdot {\hat {\bm S}}_2^{~} + 
( 1 - \gamma ) \, {\hat {\bm S}}_3^{~} \cdot {\hat {\bm S}}_4^{~}
\end{eqnarray}
shares the same ground state within the range $| \gamma | < 1$, and the
ground-state crossover 
%-ed-% happens 
occurs 
at $\gamma = \pm 1$. 

In the same manner as the $Z$-component, we can use the $X$- and 
the $Y$-components to obtain different types of modulated parts
\begin{eqnarray}
{\hat H}_X^{(4)} \!\!\!\! &=& \!\!\!\! 
{\hat {\bm S}}_2^{~} \cdot {\hat {\bm S}}_3^{~} - 
{\hat {\bm S}}_1^{~} \cdot {\hat {\bm S}}_4^{~} \, , \\
{\hat H}_Y^{(4)} \!\!\!\!  &=& \!\!\!\! 
{\hat {\bm S}}_2^{~} \cdot {\hat {\bm S}}_4^{~} - 
{\hat {\bm S}}_1^{~} \cdot {\hat {\bm S}}_3^{~} \, .
\end{eqnarray}
A simple analysis shows that the ground state of the combined Hamiltonian
\begin{equation}
{\hat H}_{~}^{(4)}( \alpha, \beta, \gamma ) = {\hat H}_0^{(4)} 
+ \alpha \, {\hat H}_X^{(4)} + \beta \, {\hat H}_Y^{(4)} + \gamma \, {\hat H}_Z^{(4)} 
\end{equation}
is independent of the coefficients $\alpha$, $\beta$, and $\gamma$ within the range~\cite{Kohmoto}
\begin{equation}
\alpha^2_{~} + \beta^2_{~} + \gamma^2_{~} \le 3 ~~~ {\rm and} ~~~
\alpha^2_{~} + \beta^2_{~} + \gamma^2_{~} + 2 \alpha \beta \gamma \le 1 \, .
\end{equation}
Typical values of the parameters at the boundary of the above region are 
$(\alpha, \beta, \gamma) = (\frac{1}{\sqrt{2}},  \frac{1}{\sqrt{2}}, 0)$, 
$(\frac{1}{2}, \frac{1}{2}, \frac{1}{2})$, and $(-1,-1,-1)$.

The analysis of the tetrahedral cluster has shown that the perturbative terms ${\hat H}_X^{(4)}$, 
${\hat H}_Y^{(4)}$, and ${\hat H}_Z^{(4)}$ do not alter the ground state of ${\hat H}_0^{(4)}$ if 
their magnitudes are sufficiently small. In the following sections, we will observe similar results 
for the ground states of larger polyhedral clusters.

\section{On the Octahedral and Cubic Clusters}

\begin{figure}
\begin{center}
\centering\includegraphics[width = 52 mm]{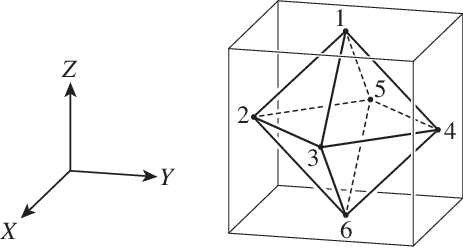}
\end{center}
\caption{Positions of sites on the octahedral cluster.}
\end{figure}

The second system we consider is the octahedral cluster shown in Fig.~3, where the
site coordinates are fixed as
\begin{align}
& 1:( \, 0, ~0, ~1 \, ) \, , ~~~ 2:( ~~0, - 1, ~0 \, ) \, , ~~~ 3:( ~1, ~0, ~~0 \, ) \nonumber\\
& 4:( \, 0, ~1, ~0 \, ) \, , ~~~ 5:( - 1, ~~0, ~0 \, ) \, , ~~~ 6:( ~0, ~0, - 1 \, ) \, .
\end{align}
In this case, the uniform Hamiltonian is given by
\begin{eqnarray}
{\hat H}_0^{(6)} \!\!\!\! &=& \!\!\!\! 
{\hat {\bm S}}_1^{~} \cdot {\hat {\bm S}}_2^{~} + 
{\hat {\bm S}}_1^{~} \cdot {\hat {\bm S}}_3^{~} + 
{\hat {\bm S}}_1^{~} \cdot {\hat {\bm S}}_4^{~} + 
{\hat {\bm S}}_1^{~} \cdot {\hat {\bm S}}_5^{~}  \nonumber \\
\!\!\!\! &+& \!\!\!\! 
{\hat {\bm S}}_2^{~} \cdot {\hat {\bm S}}_3^{~} + 
{\hat {\bm S}}_3^{~} \cdot {\hat {\bm S}}_4^{~} + 
{\hat {\bm S}}_4^{~} \cdot {\hat {\bm S}}_5^{~} + 
{\hat {\bm S}}_5^{~} \cdot {\hat {\bm S}}_2^{~}  \nonumber \\
\!\!\!\! &+& \!\!\!\! 
{\hat {\bm S}}_2^{~} \cdot {\hat {\bm S}}_6^{~} + 
{\hat {\bm S}}_3^{~} \cdot {\hat {\bm S}}_6^{~} + 
{\hat {\bm S}}_4^{~} \cdot {\hat {\bm S}}_6^{~} + 
{\hat {\bm S}}_5^{~} \cdot {\hat {\bm S}}_6^{~} \, ,
\end{eqnarray}
which has a non-degenerate ground state $| \, \psi_0^{(6)} \rangle$. 
On the lattice, the $Z$-components of the midpoints of the bonds 
are $1/2$ for $( 1, 2 )$, $( 1, 3 )$, $( 1, 4 )$, and $( 1, 5 )$, and 
are $- 1/2$ for $( 2, 6 )$, $( 3, 6 )$, $( 4, 6 )$, and $( 5, 6 )$, and 
are $0$ otherwise. To simplify the notation, we multiply the factor $2$ to these $Z$-components 
to define the modulated part
\begin{eqnarray}
{\hat H}_Z^{(6)} \!\!\!\! &=& \!\!\!\! 
{\hat {\bm S}}_1^{~} \cdot {\hat {\bm S}}_2^{~} + 
{\hat {\bm S}}_1^{~} \cdot {\hat {\bm S}}_3^{~} + 
{\hat {\bm S}}_1^{~} \cdot {\hat {\bm S}}_4^{~} + 
{\hat {\bm S}}_1^{~} \cdot {\hat {\bm S}}_5^{~} 
\nonumber\\
 \!\!\!\! &-& \!\!\!\! 
{\hat {\bm S}}_2^{~} \cdot {\hat {\bm S}}_6^{~} - 
{\hat {\bm S}}_3^{~} \cdot {\hat {\bm S}}_6^{~} - 
{\hat {\bm S}}_4^{~} \cdot {\hat {\bm S}}_6^{~} - 
{\hat {\bm S}}_5^{~} \cdot {\hat {\bm S}}_6^{~} \, .
\end{eqnarray}
%
%-ed-% Though 
Although 
${\hat H}_Z^{(6)}$ does not commute with ${\hat H}_0^{(6)}$, the relation
\begin{equation}
{\hat H}_Z^{(6)} \, | \, \psi_0^{(6)} \rangle = 0 
\end{equation}
holds, and 
%-ed-% therefore 
thus 
$| \, \psi_0^{(6)} \rangle$ is an eigenstate of ${\hat H}_Z^{(6)}$ 
with eigenvalue zero. In the same manner as we have introduced ${\hat H}_Z^{(6)}$, 
we can use the $X$- and $Y$-components of the coordinates, respectively, to define
\begin{eqnarray}
{\hat H}_X^{(6)} \!\!\!\! &=& \!\!\!\! 
{\hat {\bm S}}_3^{~} \cdot {\hat {\bm S}}_1^{~} + 
{\hat {\bm S}}_3^{~} \cdot {\hat {\bm S}}_2^{~} + 
{\hat {\bm S}}_3^{~} \cdot {\hat {\bm S}}_6^{~} + 
{\hat {\bm S}}_3^{~} \cdot {\hat {\bm S}}_4^{~} 
\nonumber\\
 \!\!\!\! &-& \!\!\!\! 
{\hat {\bm S}}_1^{~} \cdot {\hat {\bm S}}_5^{~} - 
{\hat {\bm S}}_2^{~} \cdot {\hat {\bm S}}_5^{~} - 
{\hat {\bm S}}_6^{~} \cdot {\hat {\bm S}}_5^{~} - 
{\hat {\bm S}}_4^{~} \cdot {\hat {\bm S}}_5^{~} 
\end{eqnarray}
and 
\begin{eqnarray}
{\hat H}_Y^{(6)} \!\!\!\! &=& \!\!\!\! 
{\hat {\bm S}}_4^{~} \cdot {\hat {\bm S}}_1^{~} + 
{\hat {\bm S}}_4^{~} \cdot {\hat {\bm S}}_3^{~} + 
{\hat {\bm S}}_4^{~} \cdot {\hat {\bm S}}_6^{~} + 
{\hat {\bm S}}_4^{~} \cdot {\hat {\bm S}}_5^{~} 
\nonumber\\
 \!\!\!\! &-& \!\!\!\! 
{\hat {\bm S}}_1^{~} \cdot {\hat {\bm S}}_2^{~} - 
{\hat {\bm S}}_3^{~} \cdot {\hat {\bm S}}_2^{~} - 
{\hat {\bm S}}_6^{~} \cdot {\hat {\bm S}}_2^{~} - 
{\hat {\bm S}}_5^{~} \cdot {\hat {\bm S}}_2^{~} \, ,
\end{eqnarray}
where the relations
${\hat H}_X^{(6)} \, | \, \psi_0^{(6)} \rangle = 0$ and 
${\hat H}_Y^{(6)} \, | \, \psi_0^{(6)} \rangle = 0$ are also satisfied.

\begin{figure}
\begin{center}
\centering\includegraphics[width = 74 mm]{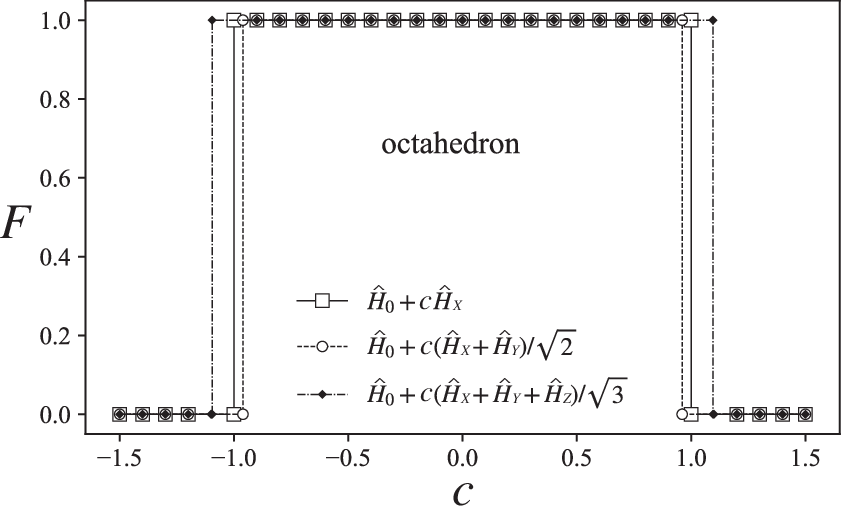}
\end{center}
\caption{Fidelity $F^{(6)}_{~}$ in the octahedral cluster.}
\end{figure}

Analogous to Eq.~(14), we introduce the linear combination 
\begin{equation}
{\hat H}^{(6)}_{~}( \alpha, \beta, \gamma ) = 
{\hat H}_0^{(6)} + 
\alpha \, {\hat H}_X^{(6)} + 
\beta \, {\hat H}_Y^{(6)} + 
\gamma \, {\hat H}_Z^{(6)} \, ,
\end{equation}
and regard it as the deformed Hamiltonian. Note that $| \, \psi_0^{(6)} \rangle$ is 
an eigenvector of ${\hat H}^{(6)}_{~}( \alpha, \beta, \gamma )$. In order to determine the
parameter region where $| \, \psi_0^{(6)} \rangle$ is the ground state of 
${\hat H}^{(6)}_{~}( \alpha, \beta, \gamma )$, we numerically diagonalize 
${\hat H}^{(6)}_{~}( \alpha, \beta, \gamma )$ to obtain its ground state 
$| \, \varphi^{(6)}_{~}( \alpha, \beta, \gamma) \rangle$, and calculate the fidelity
\begin{eqnarray}
F^{(6)}_{~}( \alpha, \beta, \gamma )  
\!\!\!\! &=& \!\!\!\! 
\left| \,  \langle \varphi^{(6)}_{~}( \alpha, \beta, \gamma ) \, | \, \psi_0^{(6)} \rangle \, 
\right|
\nonumber\\
\!\!\!\! &=& \!\!\!\! 
\left| \, \langle \varphi^{(6)}_{~}( \alpha, \beta, \gamma ) \, | \, \varphi^{(6)}_{~}( 0, 0, 0 ) \rangle \, 
\right| \, .
\end{eqnarray}
Throughout this article, we assume that all the states are normalized. 
We trace $F^{(6)}_{~}( \alpha, \beta, \gamma )$ typically 
along the paths on which the parameters are given by
\begin{eqnarray}
 &({\rm I})& \alpha = c \, , ~~~ \beta = 0 \, , ~~~ \gamma = 0 \, , \\
 &({\rm II})& \alpha = \frac{c}{\sqrt{2}} \, , ~~~ \beta = \frac{c}{\sqrt{2}} \, , ~~~ \gamma = 0 \, , \\
 &({\rm III})& \alpha = \frac{c}{\sqrt{3}} \, , ~~~ \beta = \frac{c}{\sqrt{3}} \, , ~~~ \gamma = \frac{c}{\sqrt{3}}
\, ,
\end{eqnarray}
where the factor $c$ denotes the magnitude of deformation. Figure 4 shows the 
calculated result. The fidelity $F^{(6)}_{~}$ is equal to unity for small $|c|$, and jumps to 
zero at (I) $| c | = 1$, (II) $| c | = 0.9608$, and (III) $| c | = 1.0954$, where we 
have used the parametrization in 
%-ed-% Eqs.~(24)-(26).
Eqs.~(24)--(26).
% (I) $1.0$
% (II) $0.960768432617187$ $* 2 / \sqrt{2} = 1.3586$
% (III) $1.095444946289063$ $* 2 / \sqrt{3} = 1.2644$
In cases (II) and (III), a pairwise interaction with negative coefficient appears in 
${\hat H}^{(6)}_{~}( \alpha, \beta, \gamma )$ in the neighborhood of the jumping point.

\begin{figure}
\begin{center}
\centering\includegraphics[width = 54 mm]{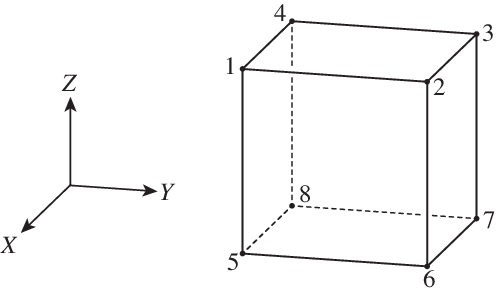}
\end{center}
\caption{Positions of sites on the cubic cluster.}
\end{figure}

The third system we consider is the cubic cluster shown in Fig.~5, where the 
uniform Hamiltonian is written as
\begin{eqnarray}
{\hat H}_0^{(8)} \!\!\!\! &=& \!\!\!\! 
{\hat {\bm S}}_1^{~} \cdot {\hat {\bm S}}_2^{~} + {\hat {\bm S}}_2^{~} \cdot {\hat {\bm S}}_3^{~} + 
{\hat {\bm S}}_3^{~} \cdot {\hat {\bm S}}_4^{~} + {\hat {\bm S}}_4^{~} \cdot {\hat {\bm S}}_1^{~} 
\nonumber\\
\!\!\!\! &+& \!\!\!\! 
{\hat {\bm S}}_1^{~} \cdot {\hat {\bm S}}_5^{~} + {\hat {\bm S}}_2^{~} \cdot {\hat {\bm S}}_6^{~} + 
{\hat {\bm S}}_3^{~} \cdot {\hat {\bm S}}_7^{~} + {\hat {\bm S}}_4^{~} \cdot {\hat {\bm S}}_8^{~} 
\nonumber\\
\!\!\!\! &+& \!\!\!\! 
{\hat {\bm S}}_5^{~} \cdot {\hat {\bm S}}_6^{~} + {\hat {\bm S}}_6^{~} \cdot {\hat {\bm S}}_7^{~} + 
{\hat {\bm S}}_7^{~} \cdot {\hat {\bm S}}_8^{~} + {\hat {\bm S}}_8^{~} \cdot {\hat {\bm S}}_5^{~} \, ,
\end{eqnarray}
which has a non-degenerate ground state $| \, \psi_0^{(8)} \rangle$. In this case, it would be 
easy to capture the lattice geometry and obtain the coordinates of the
midpoints of bonds. Similar to Eqs.~(18), (20), and (21), the modulated parts are given by
\begin{eqnarray}
{\hat H}_X^{(8)} \!\!\!\! &=& \!\!\!\! 
{\hat {\bm S}}_1^{~} \cdot {\hat {\bm S}}_2^{~} + {\hat {\bm S}}_2^{~} \cdot {\hat {\bm S}}_6^{~} + 
{\hat {\bm S}}_6^{~} \cdot {\hat {\bm S}}_5^{~} + {\hat {\bm S}}_5^{~} \cdot {\hat {\bm S}}_1^{~} 
\nonumber\\
\!\!\!\! &-& \!\!\!\! 
{\hat {\bm S}}_4^{~} \cdot {\hat {\bm S}}_3^{~} - {\hat {\bm S}}_3^{~} \cdot {\hat {\bm S}}_7^{~} - 
{\hat {\bm S}}_7^{~} \cdot {\hat {\bm S}}_8^{~} - {\hat {\bm S}}_8^{~} \cdot {\hat {\bm S}}_4^{~} \, , \\
{\hat H}_Y^{(8)} \!\!\!\! &=& \!\!\!\! 
{\hat {\bm S}}_2^{~} \cdot {\hat {\bm S}}_3^{~} + {\hat {\bm S}}_3^{~} \cdot {\hat {\bm S}}_7^{~} + 
{\hat {\bm S}}_7^{~} \cdot {\hat {\bm S}}_6^{~} + {\hat {\bm S}}_6^{~} \cdot {\hat {\bm S}}_2^{~}  
\nonumber\\
\!\!\!\! &-& \!\!\!\! 
{\hat {\bm S}}_1^{~} \cdot {\hat {\bm S}}_4^{~} - {\hat {\bm S}}_4^{~} \cdot {\hat {\bm S}}_8^{~} - 
{\hat {\bm S}}_8^{~} \cdot {\hat {\bm S}}_5^{~} - {\hat {\bm S}}_5^{~} \cdot {\hat {\bm S}}_1^{~} \, , 
\end{eqnarray}
and
\begin{eqnarray}
{\hat H}_Z^{(8)} \!\!\!\! &=& \!\!\!\! 
{\hat {\bm S}}_1^{~} \cdot {\hat {\bm S}}_2^{~} + {\hat {\bm S}}_2^{~} \cdot {\hat {\bm S}}_3^{~} + 
{\hat {\bm S}}_3^{~} \cdot {\hat {\bm S}}_4^{~} + {\hat {\bm S}}_4^{~} \cdot {\hat {\bm S}}_1^{~} 
\nonumber\\
\!\!\!\! &-& \!\!\!\! 
{\hat {\bm S}}_5^{~} \cdot {\hat {\bm S}}_6^{~} - {\hat {\bm S}}_6^{~} \cdot {\hat {\bm S}}_7^{~} -
{\hat {\bm S}}_7^{~} \cdot {\hat {\bm S}}_8^{~} - {\hat {\bm S}}_8^{~} \cdot {\hat {\bm S}}_5^{~} \, .
\end{eqnarray}
Although the modulated parts ${\hat H}_X^{(8)}$, ${\hat H}_Y^{(8)}$, and ${\hat H}_Z^{(8)}$ 
do not commute with ${\hat H}_0^{(8)}$, the relations
\begin{equation}
{\hat H}_X^{(8)} \, | \, \psi_0^{(8)} \rangle = 0 \, , ~~~
{\hat H}_Y^{(8)} \, | \, \psi_0^{(8)} \rangle = 0 \, , ~~~ {\rm and} ~~~
{\hat H}_Z^{(8)} \, | \, \psi_0^{(8)} \rangle = 0 
\end{equation}
are satisfied. 

\begin{figure}
\begin{center}
\centering\includegraphics[width = 74 mm]{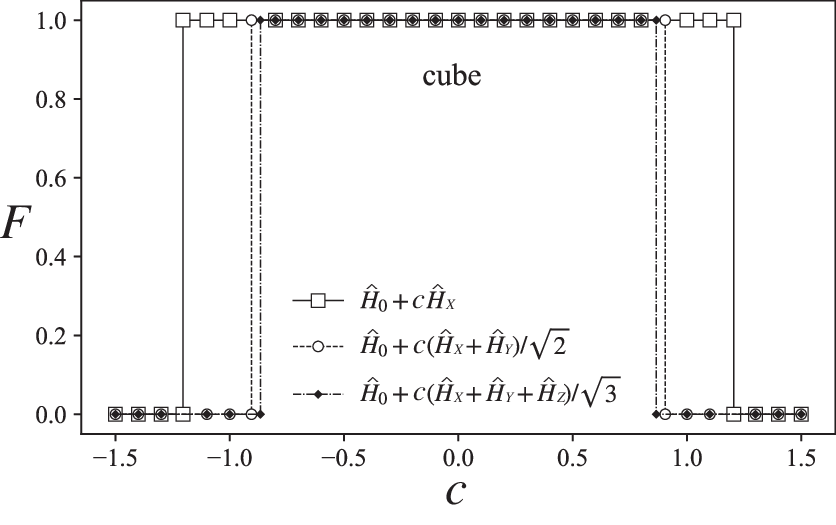}
\end{center}
\caption{Fidelity $F^{(8)}_{~}$ in the cubic cluster.}
\end{figure}

As was done in Eq.~(22), we consider the combined Hamiltonian 
${\hat H}^{(8)}_{~}( \alpha, \beta, \gamma ) = 
{\hat H}_0^{(8)} + \alpha \, {\hat H}_X^{(8)} + \beta \, {\hat H}_Y^{(8)} + \gamma \, {\hat H}_Z^{(8)}$. 
%-ed-% It should be noted 
Note that $| \, \psi_0^{(8)} \rangle$ is an eigenvector of 
${\hat H}^{(8)}_{~}( \alpha, \beta, \gamma )$. In order to determine the parameter
region where $| \, \psi_0^{(8)} \rangle$ is the ground state, we numerically 
diagonalize ${\hat H}^{(8)}_{~}( \alpha, \beta, \gamma )$ and obtain the 
corresponding ground state $| \, \varphi^{(8)}_{~}( \alpha, \beta, \gamma ) \rangle$.
Figure 6 shows the fidelity $F^{(8)}_{~}( \alpha, \beta, \gamma) = \left| \, 
\langle \varphi^{(8)}_{~}( \alpha, \beta, \gamma ) \, | \, \psi_0^{(8)} \rangle \, \right|$. 
Under the parametrization in 
%-ed-% Eqs.~(24)-(26), 
Eqs.~(24)--(26), 
the fidelity 
$F^{(8)}_{~}$ is equal to unity when $| c |$ is small, and jumps to zero at (I) $| c | = 1.2047$, 
(II) $| c | = 0.9050$, and (III) $| c | = 0.8660$, where the ground state alternates.
%
% (I) $1.204697265625000$
% (II) $0.904973144531250$ $* 2 / \sqrt{2} = 1.2799$
% (III) $0.866025390625000$ $* 2 / \sqrt{3} = 1.0000$
In cases (I) and (II), a pairwise interaction with negative coefficient appears in 
${\hat H}^{(8)}_{~}( \alpha, \beta, \gamma )$ in the neighborhood of the jumping point.

On the tetrahedral ($N = 4$), octahedral 
%-ed-% ($N = 6$) 
($N = 6$), 
and cubic ($N = 8$) clusters, we have 
confirmed that the ground state of the uniform part ${\hat H}_0^{(N)}$ is also a zero-energy 
eigenstate of the modulated parts ${\hat H}_X^{(N)}$, ${\hat H}_Y^{(N)}$, and ${\hat H}_Z^{(N)}$. 
This is the reason why the fidelity $F^{(N)}_{~}$ is unity in these 
systems, when the magnitude of modulation $|c|$ is relatively small.

\section{On the Icosahedral and Dodecahedral Clusters}

\begin{figure}
\begin{center}
\centering\includegraphics[width = 54 mm]{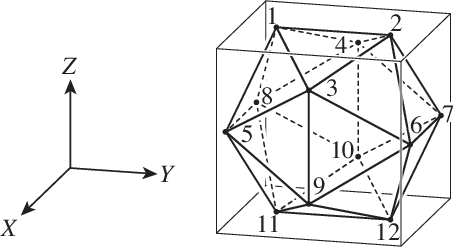}
\end{center}
\caption{Positions of sites on the icosahedral cluster.}
\end{figure}

Let us move on to the icosahedral cluster shown in Fig.~7. To save space, 
we will not write down the explicit form of
the uniform part ${\hat H}_0^{(12)}$, which is nothing but the sum of Heisenberg 
interactions ${\hat {\bm S}}_i^{~} \cdot {\hat {\bm S}}_j^{~}$ between neighboring sites. 
As we have considered in the previous sections, the modulated part is defined through the 
coordinates of the midpoints of the bonds. The golden ratio
\begin{equation}
\phi = \frac{1 + \sqrt{5}}{2} 
\end{equation}
plays an important role in writing the coordinates simply. From the $Z$-component of the 
midpoint of each bond, we obtain the modulated part
\begin{eqnarray}
H_Z^{(12)} \!\!\!\! &=& \!\!\!\! 
{\bm S}_1^{~} \cdot {\bm S}_2^{~} \nonumber\\
\!\!\!\! &+& \!\! \frac{\phi}{2} \, \Bigl(
{\bm S}_1^{~} \cdot {\bm S}_3^{~} + {\bm S}_2^{~} \cdot {\bm S}_3^{~} + 
{\bm S}_2^{~} \cdot {\bm S}_4^{~} + {\bm S}_1^{~} \cdot {\bm S}_4^{~} \Bigr) \nonumber\\
\!\!\!\! &+& \!\! \frac{1}{2} \, \Bigl(
{\bm S}_1^{~} \cdot {\bm S}_5^{~} + {\bm S}_1^{~} \cdot {\bm S}_8^{~} + 
{\bm S}_2^{~} \cdot {\bm S}_6^{~} + {\bm S}_2^{~} \cdot {\bm S}_7^{~} \Bigr) \nonumber\\
\!\!\!\! &+& \!\!\!\! \frac{1}{2 \phi} \Bigl(
{\bm S}_3^{~} \cdot {\bm S}_5^{~} + {\bm S}_3^{~} \cdot {\bm S}_6^{~} + 
{\bm S}_4^{~} \cdot {\bm S}_7^{~} + {\bm S}_4^{~} \cdot {\bm S}_8^{~} \Bigr) \nonumber\\
\!\!\!\! &-& \!\!\!\! \frac{1}{2 \phi} \Bigl(
{\bm S}_9^{~} \cdot {\bm S}_5^{~} + {\bm S}_9^{~} \cdot {\bm S}_6^{~} + 
{\bm S}_{10}^{~} \cdot {\bm S}_7^{~} + {\bm S}_{10}^{~} \cdot {\bm S}_8^{~} \Bigr) \nonumber\\
\!\!\!\! &-& \!\! \frac{1}{2} \, \Bigl(
{\bm S}_{11}^{~} \cdot {\bm S}_5^{~} + {\bm S}_{11}^{~} \cdot {\bm S}_8^{~} + 
{\bm S}_{12}^{~} \cdot {\bm S}_6^{~} + {\bm S}_{12}^{~} \cdot {\bm S}_7^{~} \Bigr) \nonumber\\
\!\!\!\! &-& \!\! \frac{\phi}{2} \, \Bigl(
{\bm S}_{11}^{~} \cdot {\bm S}_9^{~} + {\bm S}_{12}{~} \cdot {\bm S}_9^{~} + 
{\bm S}_{12}^{~} \cdot {\bm S}_{10}^{~} + {\bm S}_{11}^{~} \cdot {\bm S}_{10}^{~} \Bigr) \nonumber\\
\!\!\!\! &-& \!\!\!\! {\bm S}_{11}^{~} \cdot {\bm S}_{12}^{~} \, .
\end{eqnarray}
In the same manner, we can write down ${\hat H}_X^{(12)}$ and ${\hat H}_Y^{(12)}$, 
respectively, using the $X$- and $Y$-components. 
%-ed-% It should be noted 
Note that ${\hat H}_X^{(12)}$ 
and ${\hat H}_Y^{(12)}$ can be written just by replacing the lattice indices in Eq.~(33) 
appropriately. In this case, the ground state of ${\hat H}^{(12)}$ is not an eigenstate of 
${\hat H}_X^{(12)}$, ${\hat H}_Y^{(12)}$, nor ${\hat H}_Z^{(12)}$.

\begin{figure}
\begin{center}
\centering\includegraphics[width = 74 mm]{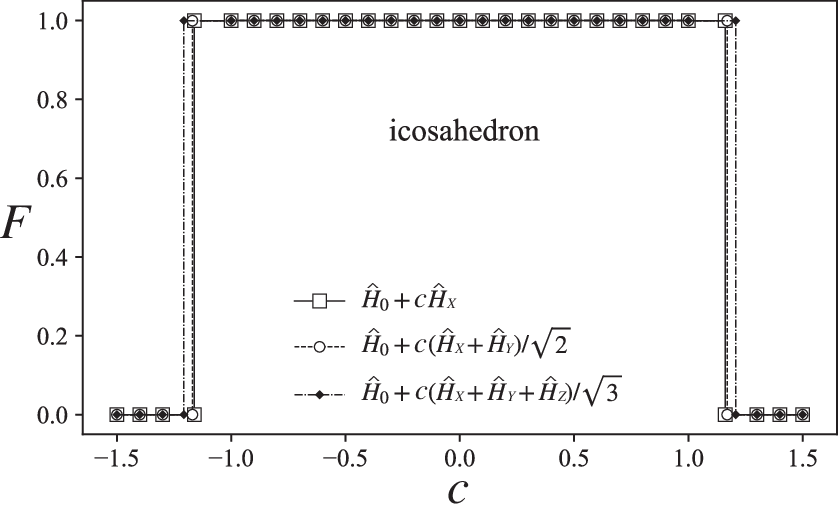}
\end{center}
\caption{Fidelity $F^{(12)}_{~}$ in the 
%-ed-% icosehedral 
icosahedral 
cluster.}
\end{figure}

As we have done in the previous section, we consider the combined 
Hamiltonian ${\hat H}_{~}^{(12)} = {\hat H}_0^{(12)} + 
\alpha \, {\hat H}_X^{(12)} + \beta \, {\hat H}_Y^{(12)} + \gamma \, {\hat H}_Z^{(12)}$, 
and obtain the ground state $| \, \varphi^{(12)}_{~}( \alpha, \beta, \gamma ) \rangle$ 
by means of numerical diagonalization. 
Figure 8 shows the fidelity $F^{(12)}_{~}( \alpha, \beta, \gamma) = \left| \, 
\langle \varphi^{(12)}_{~}( \alpha, \beta, \gamma ) \, | \, \varphi^{(12)}_{~}( 0, 0, 0 ) \rangle \, 
\right|$.
Under the parametrization in 
%-ed-% Eqs.~(24)-(26), 
Eqs.~(24)--(26),
$F^{(12)}_{~}$ is close to unity when $| c |$ is small, 
and slightly decreases with $| c |$. The values of $c$ and $F^{(12)}_{~}$ at the border where 
the fidelity changes discontinuously are (I) $| c | = 1.1613$ and $F^{(12)}_{~} = 0.9997$, 
(II) $| c | = 1.1697$ and $F^{(12)}_{~} = 0.9997$, and (III) $| c | = 1.2072$ and 
$F^{(12)}_{~} = 0.9996$. 
% (I) $1.161301879882812$  $0.999668332389357$
% (II) $1.169657592773437 > 1.08363$  $0.9996660981810526$ 
% (III) $1.207188110351562 > 1.070466$  $0.9996220859375929$
In all the cases, some of the pairwise interactions in ${\hat H}^{(12)}_{~}( \alpha, \beta, \gamma )$ 
have a negative coefficient in the neighborhood of the border.

\begin{figure}
\begin{center}
\centering\includegraphics[width = 54 mm]{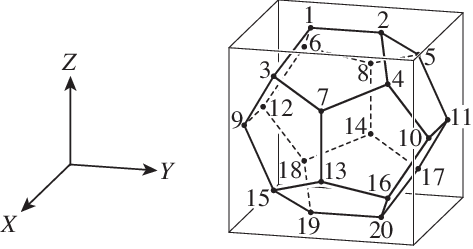}
\end{center}
\caption{Positions of sites on the dodecahedral cluster.}
\end{figure}

The last example we consider is the dodecahedral cluster shown in Fig.~9.
The uniform part ${\hat H}_0^{(20)}$ is the sum of neighboring Heisenberg interactions. 
The modulated part $H_Z^{(20)}$ is given by
\begin{eqnarray}
H_Z^{(20)} \!\!\!\! &=& \!\!\!\! 
{\bm S}_1^{~} \cdot {\bm S}_2^{~} \nonumber\\
\!\!\!\! &+& \!\! \frac{\phi}{2} \, \Bigl(
{\bm S}_1^{~} \cdot {\bm S}_3^{~} + {\bm S}_2^{~} \cdot {\bm S}_4^{~} + 
{\bm S}_2^{~} \cdot {\bm S}_5^{~} + {\bm S}_1^{~} \cdot {\bm S}_6^{~} \Bigr) \nonumber\\
\!\!\!\! &+& \!\! \frac{1}{2} \, \Bigl(
{\bm S}_3^{~} \cdot {\bm S}_7^{~} + {\bm S}_7^{~} \cdot {\bm S}_4^{~} + 
{\bm S}_5^{~} \cdot {\bm S}_8^{~} + {\bm S}_8^{~} \cdot {\bm S}_6^{~} \Bigr) \nonumber\\
\!\!\!\! &+& \!\!\!\! \frac{1}{2 \phi} \Bigl(
{\bm S}_3^{~} \cdot {\bm S}_9^{~} + {\bm S}_4^{~} \cdot {\bm S}_{10}^{~} + 
{\bm S}_5^{~} \cdot {\bm S}_{11}^{~} + {\bm S}_6^{~} \cdot {\bm S}_{12}^{~} \Bigr) \nonumber\\
\!\!\!\! &-& \!\!\!\! \frac{1}{2 \phi} \Bigl(
{\bm S}_9^{~} \cdot {\bm S}_{15}^{~} + {\bm S}_{10}^{~} \cdot {\bm S}_{16}^{~} + 
{\bm S}_{11}^{~} \cdot {\bm S}_{17}^{~} + {\bm S}_{12}^{~} \cdot {\bm S}_{18}^{~} \Bigr) \nonumber\\
\!\!\!\! &-& \!\! \frac{1}{2} \, \Bigl(
{\bm S}_{15}^{~} \cdot {\bm S}_{13}^{~} + {\bm S}_{13}^{~} \cdot {\bm S}_{16}^{~} + 
{\bm S}_{17}^{~} \cdot {\bm S}_{14}^{~} + {\bm S}_{14}^{~} \cdot {\bm S}_{18}^{~} \Bigr) \nonumber\\
\!\!\!\! &-& \!\! \frac{\phi}{2} \, \Bigl(
{\bm S}_{15}^{~} \cdot {\bm S}_{19}^{~} + {\bm S}_{16}^{~} \cdot {\bm S}_{20}^{~} + 
{\bm S}_{17}^{~} \cdot {\bm S}_{20}^{~} + {\bm S}_{18}^{~} \cdot {\bm S}_{19}^{~} \Bigr) \nonumber\\
\!\!\!\! &-& \!\!\!\! {\bm S}_{19}^{~} \cdot {\bm S}_{20}^{~} \, ,
\end{eqnarray}
where ${\hat H}_X^{(20)}$ and ${\hat H}_Y^{(20)}$ can be written in the same manner.
Also in this case, the ground state of ${\hat H}^{(20)}$ is not an eigenstate of 
${\hat H}_X^{(20)}$, ${\hat H}_Y^{(20)}$, nor ${\hat H}_Z^{(20)}$.

\begin{figure}
\begin{center}
\centering\includegraphics[width = 74 mm]{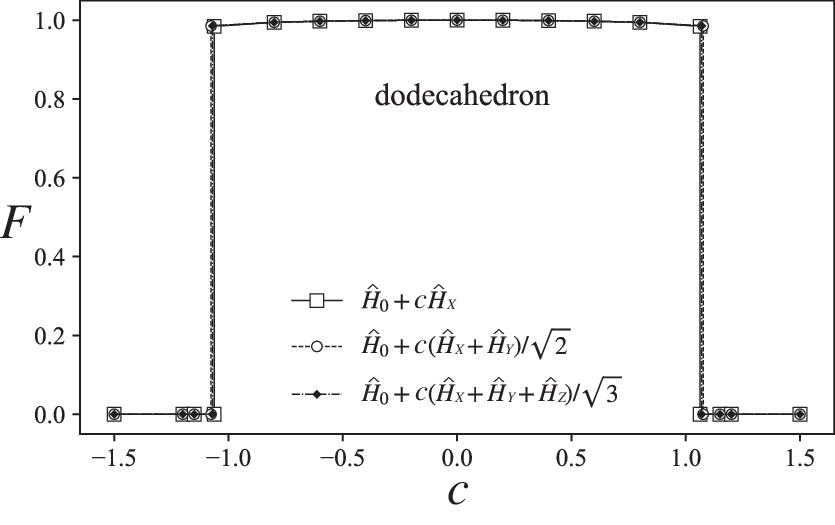}
\end{center}
\caption{Fidelity $F^{(20)}_{~}$ in the dodecahedral cluster.}
\end{figure}

We consider the combined Hamiltonian ${\hat H}_0^{(20)} + 
\alpha \, {\hat H}_X^{(20)} + \beta \, {\hat H}_Y^{(20)} + \gamma \, {\hat H}_Z^{(20)}$, 
and obtain the ground state $| \, \varphi^{(20)}_{~}( \alpha, \beta, \gamma ) \rangle$ 
by means of the numerical Lanczos diagonalization. 
Figure 10 shows the fidelity $F^{(20)}_{~}( \alpha, \beta, \gamma) = \left| \, 
\langle \varphi^{(20)}_{~}( \alpha, \beta, \gamma ) \, | \, \varphi^{(20)}_{~}( 0, 0, 0 ) \rangle \, 
\right|$.
Under the parametrization in 
%-ed-% Eqs.~(24)-(26), 
Eqs.~(24)--(26), 
$F^{(20)}_{~}$ is close to unity when $| c |$ is small, 
and slightly decreases with $| c |$. The values of $c$ and $F^{(20)}_{~}$ at the border where 
the fidelity changes discontinuously are (I) $| c | = 1.0628$ and $F^{(20)}_{~} = 0.9843$, 
(II) $| c | = 0.9843$ and $F^{(20)}_{~} = 0.9853$, and (III) $| c | = 1.07047$ and 
$F^{(20)}_{~} = 0.9857$. 
% (I) $1.062804412841797$  $0.9843172107307447$
% (II) $0.984322119140625 < 1.08363$  $0.9853033668932721$
% (III) $1.070465850830078 < 1.070466$  $0.9857144586823123$
In case (I), one of the pairwise interactions in ${\hat H}^{(20)}_{~}( \alpha, \beta, \gamma )$ 
has a negative coefficient near the border.

\section{Conclusion and Discussion}

We have examined the effect of energy scale deformation applied to the antiferromagnetic 
Heisenberg model on the polyhedral clusters. The deformation is introduced by the
perturbative Hamiltonian, which is defined through the coordinate of the midpoint of each 
bond. In the tetrahedral, octahedral, and cubic clusters, the ground states are not modified 
at all by the energy scale deformation, up to a certain amount of deformation magnitude. 
In the icosahedral and dodecahedral clusters, the ground state is slightly modified, but the 
effect of perturbation is very weak.

In our trial of the energy scale deformation, we used linear functions of the 
%-ed-% mid-point coordinate
midpoint coordinate 
of each bond as the prefactor of the modulated part. There would be a better construction of 
the modulated part also in the icosahedral and dodecahedral clusters, where the uniform 
Hamiltonian and the modulated part share a common eigenstate. Since the parameter space 
of the prefactors is finite, one way to clarify this issue is to perform a parameter search 
numerically. We expect that the symmetries of the polyhedra restrict the number of 
independent parameters, thereby making the analysis simpler. 
%%%%%-begin-%%%%%
Another possible approach is to find out the most slowly varying function on the polyhedral
lattice by means of the diagonalization of the one-particle hopping Hamiltonian on the lattice. 
%-ed-% It should be noted 
Note that the generation of an orthonormal set by diagonalization can be 
generalized to finite lattices with planar geometry, such as square lattices with rectangular 
or disk geometry,~\cite{Hotta,Nishimoto,Asano} with appropriate boundary conditions. 
The nearly 
%-ed% uniform, 
uniform
and the most slowly varying functions, respectively, may correspond
to ${\hat H}_0^{~}$ and the modulated part ${\hat H}_{\rm M}^{~}$.
%%%%%-end-%%%%%

The deformation effect can also be examined on the Archimedean solids, the quasi-regular 
polyhedra with a larger number of 
%-ed-% sites,
sites~\cite{Konstantinidis2,Coffey0,Schnack1,Exler,Schnack2,Rousochatzakis,
Schnack3,Ummethuma1,Ummethuma2,Tabrizi} such as the C60 ``buckyball".~\cite{Coffey,Modine,Rausch} 
An interesting question to ask is whether the deformation effect decreases with 
the number of sites $N$ towards the continuous limit on the sphere. 
In four dimensions, there are several regular polytope (or poly-cell) models, and the effect of 
energy scale deformation can be considered on these systems. 
%-ed-% Since the largest case is $N = 600$, and therefore one has to 
For the largest case, the 600-cell ($N = 600$), one must
employ the tensor network method to obtain the ground state.

\begin{acknowledgement}

The authors are grateful to K.~Okunishi for valuable discussions. 
%-ed-% T.N. 
T. N. 
acknowledges support from 
%-ed-% a 
JSPS KAKENHI Grant 
%-ed-% No. 
Nos. 17K05578 and 21K03403.
H. K. was supported by JSPS Grant-in-Aid for Scientific Research on Innovative Areas No. JP20H04630, JSPS KAKENHI Grant No. JP18K03445, and the Inamori Foundation.

\end{acknowledgement}

\end{document}